\documentclass[prr,aps,twocolumn,superscriptaddress]{revtex4-2}

\usepackage[none]{hyphenat}
\usepackage{amsmath}
\usepackage{amssymb}
\usepackage{color}
\usepackage{feynmp-auto}
\usepackage{graphicx}
\usepackage{grffile}
\usepackage{placeins}
\usepackage[colorlinks=true,citecolor=blue,linkcolor=blue]{hyperref}
\usepackage{mathtools}
\graphicspath{{images/}}
\graphicspath{{Figures/}}

\newcommand{\avg}[1]{\left<#1\right>} 
\newcommand{\ket}[1]{\big| #1 \big>} 

\begin{document}
\title{Dynamically generated quadrupole polarization using Floquet adiabatic evolution}

\author{G.~Camacho}
\affiliation{Technische Universit\"at Braunschweig, Institut f\"ur Mathematische Physik, Mendelssohnstrasse 3, 38106 Braunschweig, Germany}
\author{C.~Karrasch}
\affiliation{Technische Universit\"at Braunschweig, Institut f\"ur Mathematische Physik, Mendelssohnstrasse 3, 38106 Braunschweig, Germany}

\author{R.~Rausch}
\affiliation{Technische Universit\"at Braunschweig, Institut f\"ur Mathematische Physik, Mendelssohnstrasse 3, 38106 Braunschweig, Germany}

\setlength{\abovedisplayskip}{4pt}
\setlength{\belowdisplayskip}{4pt}  


\begin{abstract}
We investigate the nonequilibrium dynamics of the $S=1$ quantum spin chain subjected to a time-dependent external drive, where the driving frequency is adiabatically decreased as a function of time (``Floquet adiabatic evolution''). We show that when driving the rhombic anisotropy term (known as the ``two-axis countertwisting'' in the context of squeezed spin states) of a Néel antiferromagnet, we are able to induce an overall enhancement in the quadrupole polarization, while at the same time suppressing the staggered magnetization order. The system evolves into a new state with a net quadrupole moment and antiferroquadrupolar correlations. This state remains stable at long times once the driving frequency is kept constant. On the other hand, we find that we cannot achieve a quadrupole polarization for the symmetry-protected Haldane phase, which remains robust against such driving.
\end{abstract}

\maketitle

\section{Introduction}

In order to find quantum states with desired properties, we can look in various spaces: In the chemical space we can investigate the multitude of natural compounds. This space can be further extended by synthesis, metamaterials or the replacement of chemical bonds by magneto-optical traps in ultracold quantum gases. Another possibility is to exploit the additional dimension of time and engineer new states in nonequilibrium conditions.

A particularly simple way to create a nonequilibrium state is a quantum quench, where the system is suddenly evolved with a new Hamiltonian. This can be used to observe the melting of equilibrium order parameters, such as string order~\cite{Mazza_2014,CalvaneseStrinati_2016}, and can lead to quasi-steady prethermalized states~\cite{Moeckel_2008} before heating sets in, but does not offer optimal control and is not easily implementable beyond ultracold-atom systems.

Alternatively, one can drive the system out of equilibrium by a periodic external force, e.g., a continuous laser beam with frequency $\Omega=2\pi/T$ and period $T$. In practical terms, this setup is well-controlled in the high-frequency limit, where one can find the effective \emph{Floquet Hamiltonian}~\cite{floquet_analytic_review, Bukov_review} by means of a Magnus expansion of the original Schr{\"o}dinger equation. In leading order, this results in renormalized system parameters, so that the problem can be analyzed using equilibrium techniques (\emph{Floquet engineering}). Heating to an infinite-temperature state should take place eventually, but is shown to happen on exponentially long time scales for large frequencies, leading to a stable prethermalized state similar to the case of quenches~\cite{Abanin_2015,Kitamura_2016}. 

Floquet engineering has been applied very extensively~\cite{Oka_2019,Weitenberg2021,RevModPhysNonthermal2021,Kennes2018} and a comprehensive listing of all results is near-impossible. For non-interacting systems, the electronic band structure is modified, which becomes interesting if the topological character is changed~\cite{Oka_2009,Kitagawa_2010,Lindner_2011,Rudner_2013}. For interacting systems, a lot of attention has been devoted to the 
enhancement of superconducting correlations~\cite{Fausti2011,Mitrano2016,Babadi_2017,Sheikhan2020} (often using intense pulses rather than continuous beams) and the photo-inducement of superconducting orders absent from equilibrium phases, such as $\eta$-pairing~\cite{Kitamura_2016,Kaneko2019,Peronaci_2020}. Apart from that, there have been efforts to control the Kondo effect~\cite{Takasan_2017}, exchange interactions~\cite{Mentink_2015}, the Dzyaloshinskii-Moriya interaction~\cite{Sato2016}, the magnetization~\cite{Takayoshi}, or many-body localization~\cite{Decker2020}.

In equilibrium physics, the concept of adiabaticity is fundamental. In practical terms, it can be used to define and traverse phase diagrams or prepare complex ground states by adiabatically changing the couplings of a Hamiltonian, e.g. using quantum annealing. Extending this concept to Floquet engineering, one can attempt to adiabatically change the drive parameters to further improve the degree of dynamic control of the system~\cite{weinberg_fpt,Guerin_2003}.

In this work, we adopt the specific protocol of initiating the system by driving a term with $\Omega=\infty$, followed by an adiabatic decrease of $\Omega$~\cite{Russomanno_KZ,Russomanno_2}. This adiabatically propagated state is called the ``Floquet ground state''~\cite{Russomanno_2}, and $\Omega$ is freed up as an additional control parameter in the procedure.
This has been first studied for the integrable transverse-field Ising model, in which case the state was seen to undergo topological phase transitions and Kibble-Zurek scaling was observed~\cite{Russomanno_KZ,Russomanno_2}.

Fortunately, in the case of one-dimensional chains, this ``adiabatic Floquet'' protocol lends itself to an efficient simulation even for non-integrable systems using matrix-product states (MPS). The initial state is guaranteed to have low entanglement for the class of gapped chains in accordance with the area law. This is a key property that is exploited by the MPS formalism~\cite{Schollwock}. Furthermore, as long as the change of frequency is slow enough, the entanglement is expected to grow only slowly and long propagation times may be reached.
This stands in contrast to quench dynamics, where the entanglement entropy increases linearly with time~\cite{calabrese_cardy}, while the MPS bond dimension (i.e., the number of variational parameters to represent the state) has to increase exponentially.

In this paper, we show that the adiabatic Floquet protocol can be used to convert a conventional N\'eel antiferromagnetic state into an unconventional antiferroquadrupolar state.
More specifically, we apply the protocol to the non-integrable $S=1$ spin chain, a system which is mainly interesting for its symmetry-protected ``topological'' Haldane phase and the potential for spin-nematic (quadrupolar) order. The latter is a state with nonvanishing anisotropic second-order expectation values of the type $\avg{S^{\alpha}_jS^{\beta}_j}\neq0$, while having a vanishing first-order expectation $\avg{S_j^{\alpha}}=0$ (where $S_j^{\alpha=x,y,z}$ is a spin operator). Thus, it is an interesting quantum state that carries no magnetic moment, but still breaks the spin-rotational symmetry via a more complicated order parameter.
A spin nematic can be regarded as something between a ferromagnet and a spin liquid: While it lacks magnetic order like the latter, it still breaks the rotational symmetry like the former and has a preferred axis. The name derives from the physics of nematic liquid crystals, which in a similar sense constitute a phase between a liquid and a solid~\footnote{The correspondence arises from the comparison of quantum spin-spin correlations to classical density-density correlations, which are long-ranged in a solid (corresponding to a ferromagnet) and exponentially decaying in a liquid (corresponding to a disordered spin liquid). In a nematic liquid crystal, they are disordered like in a liquid, but rod-like molecules order along a preferred axis.}.
While quadrupolar order is in principle possible in the $S=1$ chain, in the following section we discuss that it is not easily achievable in equilibrium, motivating an extension to driven systems.

Starting from the ground state of an initial Hamiltonian with $\Omega=\infty$, we slowly drive the system from the high-frequency to the mid-frequency region, representing the wavefunction as a MPS. We find that if the initial state is in the symmetry-protected Haldane phase, it still remains remarkably robust against the drive and no new phase transition is found.
On the other hand, if the initial state is in a trivial ordered phase, then we are able to induce an overall quadrupolar moment and enhanced correlations, eventually reaching a stable phase with long-range antiferroquadrupolar order, where the staggered magnetization is suppressed.

\section{Model}

\subsection{Initial Hamiltonians}

\begin{figure}[t!]
\includegraphics[scale=0.65]{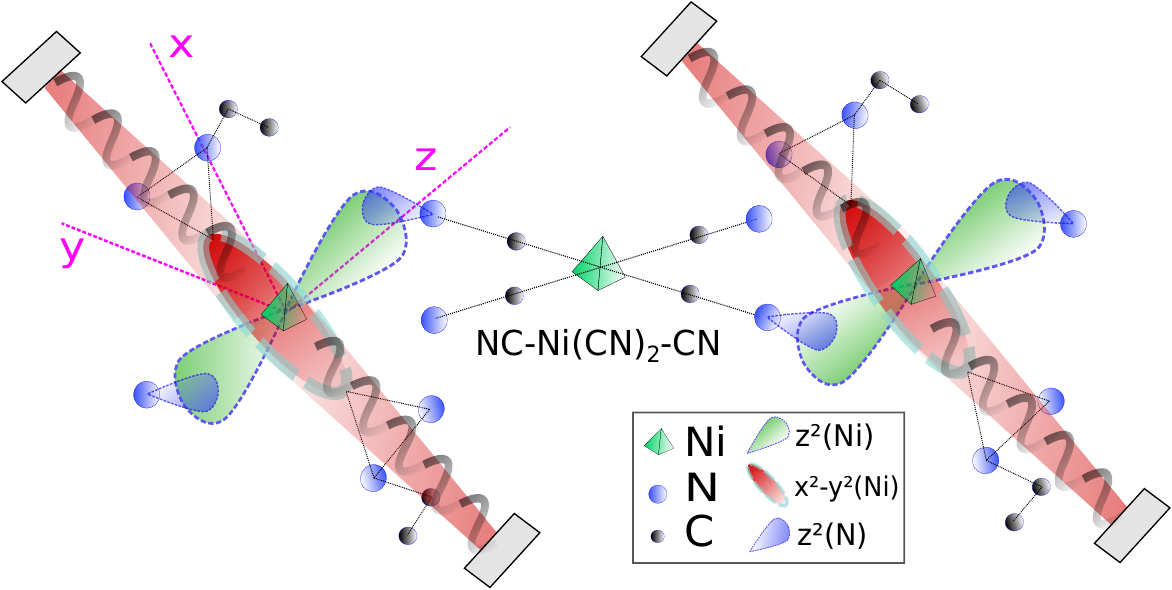}
\caption{
The compound Ni(C$_2$H$_8$N$_2$)$_2$Ni(CN)$_4$ (NENC) under the action of a periodic driving in the $x^{2}-y^{2}$ orbital of the Ni(II) atoms. The figure is inspired by Fig.~1 in Ref.~\cite{Orendac_1995}, but we have added a possible experimental driving setup. The structure of the chain repeats in the horizontal direction. Molecular orbitals from the Ni and N atoms have been represented in the $x,y,z$ geometry by the lobes (see legend). Each Ni(II) is attached to the next one by a NC-Ni(CN)$_{2}$-CN configuration. The magnetic properties of such Ni compounds have been studied experimentally. A theoretical description of these compounds is proposed by effective Hamiltonians in the form of Eq.~\eqref{Heisenberg_hamiltonian} representing the $S=1$ chain of the $\text{Ni(II)}$. The experimental realization sketched in this figure corresponds to the driving protocol given by Eq.~\eqref{protocol_XY} of this work, where the drive in the $xy$-plane (represented as a sinusoidal wave) continuously changes the probability distribution of electronic $x^{2}-y^{2}$ orbital in the Ni(II) atoms.}
\label{fig:figure_model}
\end{figure}

We consider a one-dimensional chain of localized spins with $S=1$ at zero temperature. The real system under consideration might in fact be a two- or three-dimensional array of such chains, where the interchain coupling is captured on the mean-field level by a staggered magnetic field $h$~\cite{Sakai_Takahashi_1990,Tsukano_1998,Wierschem_2014,Bera_2015}. Experimentally, such systems are realized in various Ni- and V-based compounds~\cite{Buyers_1986,Morra_1988,HaldaneXp1,HaldaneXp2,HaldaneXp3,HaldaneXp4,HaldaneXp5,HaldaneXp6,HaldaneXp7,HaldaneXp8,HaldaneXp9,HaldaneXp10,HaldaneXp11,HaldaneXp12,HaldaneXp13,HaldaneXp14,HaldaneXp15,HaldaneXp16,Lemmens_2021} (see also Fig.~\ref{fig:figure_model}).

The Hamiltonian is an extended variant of the Heisenberg spin chain:
\begin{multline}\label{Heisenberg_hamiltonian}
H_{\text{Heis}}=J\sum_{j}\vec{S}_{j}\cdot\vec{S}_{j+1} + D\sum_{j}\left(S_{j}^{z}\right)^{2}
-h\sum_{j}(-1)^{j}S_{j}^{z},
\end{multline}
where $J$ is the exchange interaction parameter. $\vec{S}_{j}=(S_{j}^{x},S_{j}^{y},S_{j}^{z})$ represents the spin-1 operator at the $j$-th site, with $S_j^{\alpha=x,y,z}$ representing the different spin projections. As the local basis $|\sigma\rangle$, we take the eigenbasis of $S_j^z$ and denote the eigenvectors as $|\sigma\rangle=\ket{+},\ket{0},\ket{-}$ with the eigenvalues of $+1$, $0$ and $-1$, respectively.
Finally, $h$ is the staggered magnetic field, while $D$ is the anisotropy in the $z$-direction (easy-axis for $D<0$ and easy-plane for $D>0$). We set $\hbar=1$ and $J=1$, thereby measuring all energies in units of $J$ and times in units of $\hbar/J$.

For $D=h=0$, there exists a gapped phase (the ``Haldane phase''), which (for periodic boundary conditions) has a unique symmetry-protected ground state with exponentially decreasing spin-spin correlations. The robustness of the Haldane phase has been a focal point of previous studies in equilibrium~\cite{Haldane_1,Haldane_2,AKLT,AKLT_2,Pollmann_2010,Pollmann_2012A,Pollmann_2012B}, where it was found that it can be characterized by a non-local string order parameter~\cite{Nijs_Rommelse, Kennedy_Tasaki, Oshikawa}
\begin{eqnarray}\label{string_order_parameter}
O^{\alpha=x,y,z}_{\text{string}}=\lim_{|j-k|\to \infty}\langle S_{j}^{\alpha}e^{\mathrm{i}\pi\sum_{l=j+1}^{k-1}S_{l}^{\alpha}} S_{k}^{\alpha}\rangle,
\end{eqnarray}
and by a global twofold degeneracy in the entanglement spectrum. It is protected by a combination of inversion symmetry, time-reversal (in the sense $S^{x,y,z}\to-S^{x,y,z}$) and combined rotations of $\pi$ about a pair of axes~\cite{Pollmann_2012A}. Since a finite $h$ breaks all these symmetries at once, even a small value destroys the Haldane phase~\cite{Tsukano_1998}. However, it remains robust against the anisotropy term, which does not break any of the above symmetries~\cite{Pollmann_2012A,Pollmann_2012B,Pollmann_2010}. In this case, the Haldane phase is a thermodynamic phase, which is stable in an extended region of the phase diagram, eventually losing in competition to strong-$D$ phases (see below) once $D$ exceeds a critical value. An interesting question is thus how this robustness extends into non-equilibrium.

The other limiting cases of Eq.~\eqref{Heisenberg_hamiltonian} are as follows: For $D\to+\infty$, the ground state is given by a product state of local $\ket{0}$ projections. In the $D\to-\infty$ limit, the ground state is two-fold degenerate, given by the N{\'e}el state $\ldots\ket{+}\ket{-}\ket{+}\ket{-}\ldots$ and the N{\'e}el state shifted by one lattice site: $\ldots\ket{-}\ket{+}\ket{-}\ket{+}\ldots$.
For $h\to\pm\infty$, the ground state is given by a unique N{\'e}el state.

The $S=1$ chain is also arguably the simplest system that allows for quadrupolar exchange. The quadrupole operator is defined as the traceless tensor
\begin{equation}
Q_j^{\alpha\beta} = S_j^{\alpha}S_j^{\beta} + S_j^{\beta}S_j^{\alpha} - \frac{2}{3}S(S+1) \delta_{\alpha\beta}.
\end{equation}
It has five linearly independent components that can be grouped into a vector:
\begin{equation}
\vec{Q}_{j}=\begin{pmatrix}
Q_j^{x^2-y^2}\\
Q_j^{3z^2-r^2}\\
Q_j^{xy}\\
Q_j^{yz}\\
Q_j^{xz}
\label{eq:Q}
\end{pmatrix}
=
\begin{pmatrix}
(S_j^x)^2-(S_j^y)^2\\
\frac{1}{\sqrt{3}} \left[3(S_j^z)^2-S(S+1)\right]\\
S_j^xS_j^y+S_j^yS_j^x\\
S_j^yS_j^z+S_j^zS_j^y\\
S_j^xS_j^z+S_j^zS_j^x
\end{pmatrix},
\end{equation}
so that $\sum_{\alpha\beta} Q_j^{\alpha\beta} Q_j^{\alpha\beta} = 2 \vec{Q}_j\cdot\vec{Q}_j$. Quadrupolar exchange thus requires a product of four spin operators and is usually discussed within the \emph{bilinear-biquadratic model}~\cite{Blume_Hsieh_1969,Barber_Batchelor_1989}, given by:
\begin{eqnarray}\label{bi_quad}
H_{\text{blbq}}= J\sum_{j}\vec{S}_{j}\cdot\vec{S}_{j+1} + J_q\sum_{j}\left(\vec{S}_{j}\cdot\vec{S}_{j+1}\right)^{2}.
\end{eqnarray}
Because of the identity
\begin{equation}
\vec{Q}_i\cdot\vec{Q}_j = 2(\vec{S}_i\cdot\vec{S}_j)^2 + \vec{S}_i\cdot\vec{S}_j - \frac{2}{3}\left[S(S+1)\right]^2,
\end{equation}
the Hamiltonian Eq.~\eqref{bi_quad} boils down to a competition of ordinary exchange interaction and quadrupolar exchange. In 1D, quantum fluctuations generally prevent spontaneous ordering unless the order parameter is conserved. In this case, a quadrupolar state may rather be defined via quadrupolar correlations that dominate over spin-spin correlations. Such correlations are found with a 3-site period for $J,J_q>0$ and $J_q/J>1$~\cite{Laeuchli_2006}. For $J<0$, a ferroquadrupolar order was initially predicted close to the ferromagnetic phase~\cite{Chubukov}, but highly accurate MPS calculations demonstrate that it either does not exist (with a dimerized phase found instead), or only exists in a very narrow parameter regime~\cite{Buchta_2005,Laeuchli_2006}.
In 2D, quadrupolar phases are better defined and a finite quadrupole moment generally arises for $\big|J_q\big|/\big|J\big|>1$~\cite{Niesen2017,Harada2002}.

We note that the choice
\begin{eqnarray}\label{aklt_hamiltonian}
H_{\text{AKLT}}=\sum_{j}\vec{S}_{j}\cdot\vec{S}_{j+1} +\frac{1}{3} \sum_{j}\left(\vec{S}_{j}\cdot\vec{S}_{j+1}\right)^{2}
\end{eqnarray}
yields the famous Affleck-Kennedy-Lieb-Tasaki (AKLT) ground state~\cite{AKLT,AKLT_2}, which belongs to the Haldane phase, but is exactly representable by an MPS with very low entanglement. From a technical point of view, it is thus a more convenient representative member of the Haldane phase than Eq.~\eqref{Heisenberg_hamiltonian} with $h=D=0$.

A possibility to generate finite quadrupolar moments in one dimension is the explicit breaking of the spin-SU(2) symmetry via the so-called ``rhombic single-ion anisotropy''~\cite{Tzeng_2017,Batchelor_2004,abragam2012electron}:
\begin{eqnarray}\label{eq:rhombic_an}
\delta H_E 
&=& E\sum_{j}\left[\left(S_{j}^{x}\right)^{2} - \left(S_{j}^{y}\right)^{2}\right]\nonumber\\
&=& E \sum_{j} Q^{x^2-y^2}_{j}
= \frac{E}{2} \sum_{j}\left[\left(S_{j}^{+}\right)^{2} + \left(S_{j}^{-}\right)^{2}\right],
\end{eqnarray}
where we have introduced the standard spin-flip operators $S_{j}^{\pm} = S^x_j\pm \mathrm{i}S^y_j$. A more general coupling to the square of the spin operator is also possible~\cite{Seifert2022,abragam2012electron}. In the context of squeezed spin states, the term in Eq.~\eqref{eq:rhombic_an} is known as the ``two-axis countertwisting'' (TACT)~\cite{Kitagawa_1993} and there are several proposals of how to implement it~\cite{Ma_2011}. In equilibrium, a strong value of $E$ will induce a finite quadrupole moment in the direction of $\avg{Q^{x^2-y^2}}$, similar to how an external field induces a finite magnetization~\cite{Tzeng_2017}. However, this requires finding a material with large $E$ and small $D$ at the same time.

Recently, attention has shifted to a different regime, where bond-nematic rather than local order might be found. For an $S=1/2$ system in a strong magnetic field close to saturation, deviations in magnetization are given by magnons. If the effective interaction between these magnons is attractive, they may form pairs and condense, with non-zero quadrupolar correlations $\avg{S^+_iS^+_j}$ playing the same role as anomalous expectation values of fermion-pair creation operators $\avg{c^{\dagger}_ic^{\dagger}_j}$ in a superconductor~\cite{Zhitomirsky2010}. The best-documented experimental example where this may occur LiCuVO$_4$~\cite{MagnonCondensate1,MagnonCondensate2}. The corresponding experiments are quite challenging and need to be performed in magnetic fields of 45-50T.

Summarizing, a spin-nematic state is an exotic nonmagnetic state with a higher-level order parameter. The minimal spin value to observe it locally is $S=1$. A stabilization of this state requires strong biquadratic exchange or a strong anisotropy. However, both are expected to be weak in real materials or require finding a fine-tuned point~\cite{Lou2000,Millet1999,Mila2000}, so that the part of the phase diagram where spin-nematic phases are predicted could not be explored in practice.
Attention has therefore shifted to the different physical regime of magnon pairing in high magnetic fields for $S=1/2$ materials, which has its own challenges.
Here, we pursue an alternative idea, namely the purposeful enhancement of quadrupolar interactions using nonequilibrium driving, starting from an ordinary $S=1$ system (Haldane chain or Néel antiferromagnet).

\subsection{Floquet adiabatic protocol}
\label{subsec:adiabatic_protocol}

We evolve a system in the presence of a time-dependent term $Af(t,\Omega(t))V$:
\begin{eqnarray}\label{ham_quench}
H(t) = H_{0} + Af(t,\Omega(t))V,
\end{eqnarray}
where $H_0$ is the unperturbed Hamiltonian, $f(t,\Omega(t))$ is the periodic envelope function of the drive, $\Omega(t)$ is the drive frequency, $A$ is its amplitude and $V$ is the operator of the drive.

It is convenient to choose an envelope function that averages to zero over one cycle,
\begin{eqnarray}\label{generic_drive}
\frac{1}{T}\int_{0}^{T}d\tau f(\tau,\Omega) = 0.
\end{eqnarray}
Specifically, we set
\begin{eqnarray}\label{drive_ft}
f(t,\Omega(t)) = \sin(\Omega(t)t).
\end{eqnarray}
This means that for $\Omega(t=0)=\infty$, the state at $t=0$ can be obtained as the ground state of $H_0$ and is a legitimate Floquet state.

In order to observe quadrupolar order in our setup, we choose to drive the rhombic anisotropy Eq.~\eqref{eq:rhombic_an}, which is quadratic in the spin operators (cf. Fig.~\ref{fig:figure_model}):
\begin{eqnarray}\label{protocol_XY}
Af(t,\Omega(t))V = Af(t,\Omega(t))\sum_{j}\left[\left(S_{j}^{x}\right)^{2} - \left(S_{j}^{y}\right)^{2}\right].
\end{eqnarray}

For $t>0$, we start with a frequency of $\Omega_0$ that is large enough to be connected with the infinite-frequency state at $t=0$. We then adiabatically decrease the frequency in the range $\Omega(t)\in[ \Omega_{f},\Omega_{0}]$ ($\Omega_{0}>\Omega_{f}$) on a time window of length $t_{f}$ (see App.~\ref{app:time_discretization} for more details). After that, we assess whether a steady state has been reached by evolving the state with constant $\Omega_{f}$ for another multiple of $t_{f}$, i.e., until the end time $t_{\text{end}}=\alpha t_f$, where $\alpha>1$ is a scaling factor. In other words, the steady state is observed on a timescale of $\left(\alpha-1\right)t_f$.

The meaning of these control parameters is as follows: (1) The target time $t_{f}$ controls adiabaticity, whereby larger values make the process more adiabatic, so that we would like to choose $t_{f}$ as large as possible.
(2) For the final frequency $\Omega_{f}$, the most interesting regime is on the energy scale of the system or below it, i.e. $\mathcal{O}(10^0)-\mathcal{O}(10^{-1})$ in units of $J$. However, if $\Omega$ becomes so small that $\Omega^{-1} \sim t_{f}$, the system cannot be considered locally periodic at any time, and the Floquet theorem loses its validity. Therefore, one should choose a range of $\Omega$ for which all values satisfy $\Omega  t_{f}\gg 2\pi$ \hspace{5pt}$\forall\Omega\in[\Omega_{f},\Omega_0]$~\cite{Russomanno_KZ}, which means that smaller $\Omega_{f}$ have to be offset by larger $t_f$.

Taking all these trade-offs into account, we set $\Omega_{0}=100.0$, $\Omega_{f}\geq 10.0$, $t_f=62.82$, $\alpha=1.5,2.5$.

\begin{figure}[t!]
\includegraphics[scale=0.56]{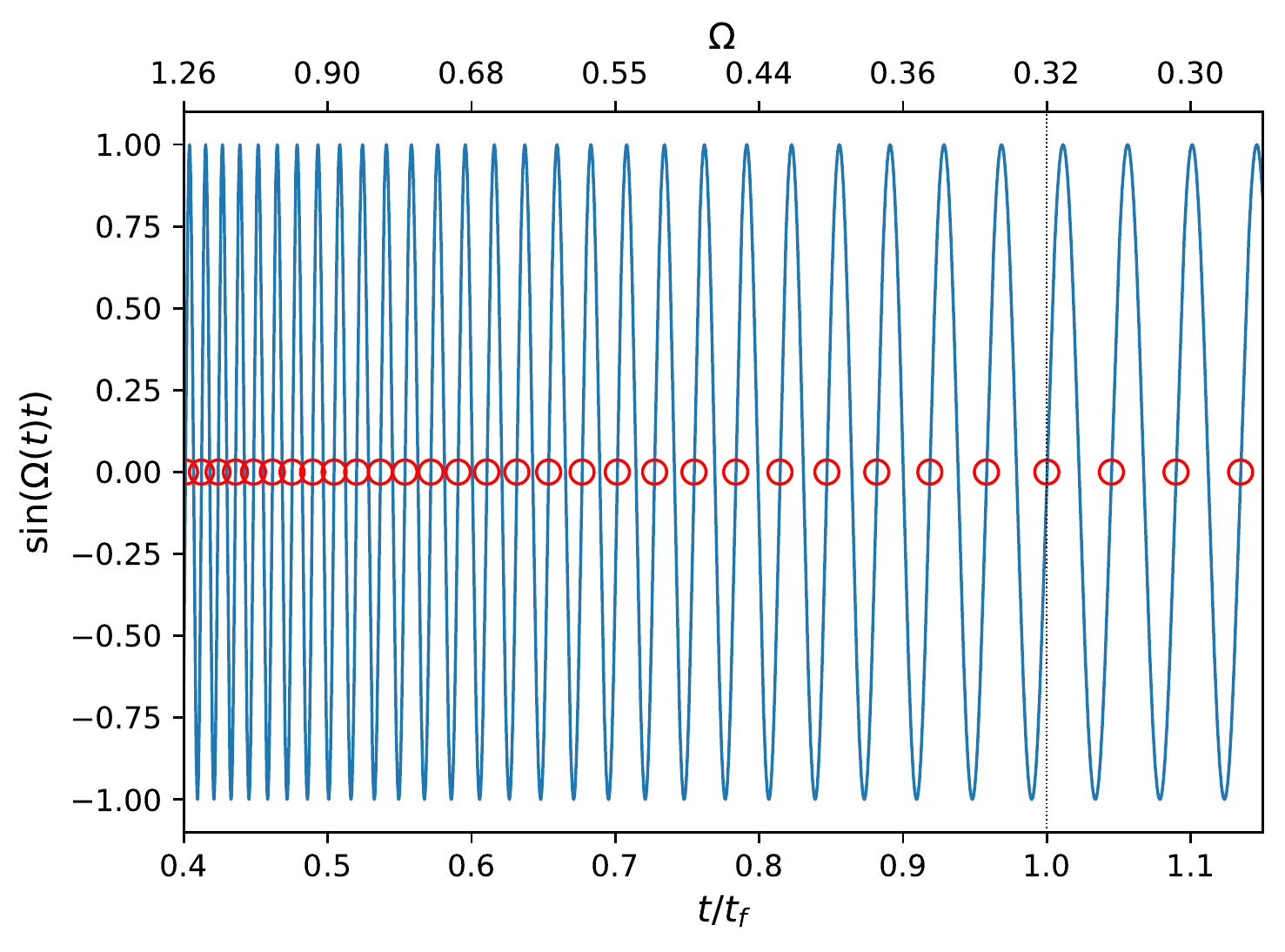}
\caption{The drive envelope function $f(t)$, Eq.~\eqref{drive_ft}, with $\Omega_{f}=0.3$. We use an \emph{intermittent} measurement process, where the red circles correspond to the measurement times of the observables. On the top axis, we have represented the corresponding values of the driving frequency $\Omega$, which is kept constant at $\Omega=\Omega_{f}$ for $t/t_{f}>1$.}
\label{figure_0}
\end{figure}

\section{Methods}
\label{sec:methods}

\subsection{Time-Evolving Block Decimation (TEBD)}
\label{subsec:tebd}

We use the \emph{infinite time-evolving block decimation} (iTEBD) with a fourth-order Suzuki-Trotter decomposition~\cite{Schollwock, time_evol_mps} to compute time evolution of a given MPS directly in the thermodynamic limit. The main control parameter is the bond dimension $\chi$, which encodes the amount of entanglement.

The MPS $|\psi\rangle$ has a unit cell of $L_{\text{cell}}=2$ for a finite staggered field in Eq.~\eqref{Heisenberg_hamiltonian}. We represent it in the standard $\Gamma-\Lambda$ notation~\cite{Vidal_tebd_1},
\begin{eqnarray}\label{eq:mps_laga}
|\psi\rangle =\sum_{\{\sigma_{i}\}}\ldots \Gamma^{\mathrm{A}\sigma_{i}}\Lambda^{\text{AB}}\Gamma^{\mathrm{B}\sigma_{i+1}}\Lambda^{\text{BA}}\ldots|\ldots\sigma_{i}\sigma_{i+1}\ldots\rangle,
\end{eqnarray}
with A and B denoting the sublattices, and $\sigma_{i}=\{+,0,-\}$. The AKLT state (Sec.~\ref{sec:Haldane}) can be expressed analytically as a MPS with a bond dimension of $\chi=2$. The N\'eel ground state (Sec.~\ref{subsec:driving_trivial_phase}) is computed from an imaginary time evolution with a fixed bond dimension of $\chi=200$ and a decreasing Trotter step size until convergence has been reached; we have checked that using $\chi=120$ or $L_{\text{cell}}=4,6$ does not change the results, and also cross-checked with the Variational Uniform MPS (VUMPS) algorithm~\cite{VUMPS} as an independent benchmark.

During the real time evolution, it is convenient to choose a variable Trotter step size $\Delta t$ that depends on the current period. We do not let it exceed the maximum value of $\Delta t=0.01$. Furthermore, we set an initially fixed bond dimension of $\chi=200$ (for the quench starting from the AKLT state, the initial bond dimension $\chi=2$ is increased such that that the state is encoded exactly until $\chi=200$ is reached). This is sufficient for an upper bound of the discarded weight (defined as the sum of discarded weights on each Trotter substep) of $\epsilon=10^{-5}$. In other words, fixing $\epsilon=10^{-5}$ does not lead to an increase of the bond dimension beyond $\chi=200$, which seems reasonable due to the adiabatic nature of the quench. Further comments on this issue as well as a comparison with exact diagonalization data can be found in App.~\ref{app:tests}.

\subsection{\label{sec:measurement}Measurement events and the steady-state limit}

When we keep track of all observables in a \emph{continuous} fashion, the micromotion of the Hamiltonian dynamics leads to the appearance of oscillations as a consequence of the drive given by Eq.~\eqref{drive_ft}. We ignore this micromotion by an \emph{intermittent} measuring at some of the zeros of the drive, as indicated in Fig.~\ref{figure_0}. This subset of $N$ measuring points is defined by:
\begin{eqnarray}\label{eq:measuring_times}
t^{*}=\{t^{*}_{j}\}\subset \left[0,\alpha t_{f}\right]\hspace{5pt}|\hspace{5pt}f(t^{*}_{j})=0\hspace{5pt}\forall j=1,...,N.
\end{eqnarray} 

Furthermore, we define the steady-state limit (SSL) value of an observable $O$ in the Heisenberg picture by averaging over the last $N_{\text{SSL}}$ measurement times:
\begin{eqnarray}\label{eq:ssl_def}
\langle O\rangle_{\text{SSL}}=\frac{\sum_{k=1}^{N_{\text{SSL}}}\langle O(t_{k}^{*})\rangle}{N_{\text{SSL}}},\hspace{10pt}t_{k}^{*}>2t_{f}.
\end{eqnarray}
Due to computational limitations, the above definition is strictly speaking a \emph{quasi}-steady state, in the sense that the reachable times are finite, but long on the intrinsic time scale: $\alpha t_{f}\gg \hbar/J = 1$ ($\alpha=1.5,2.5$).

\section{Results}

\subsection{\label{sec:Haldane}Driving from the Haldane phase}

\begin{figure}[t!]
\includegraphics[scale=0.56]{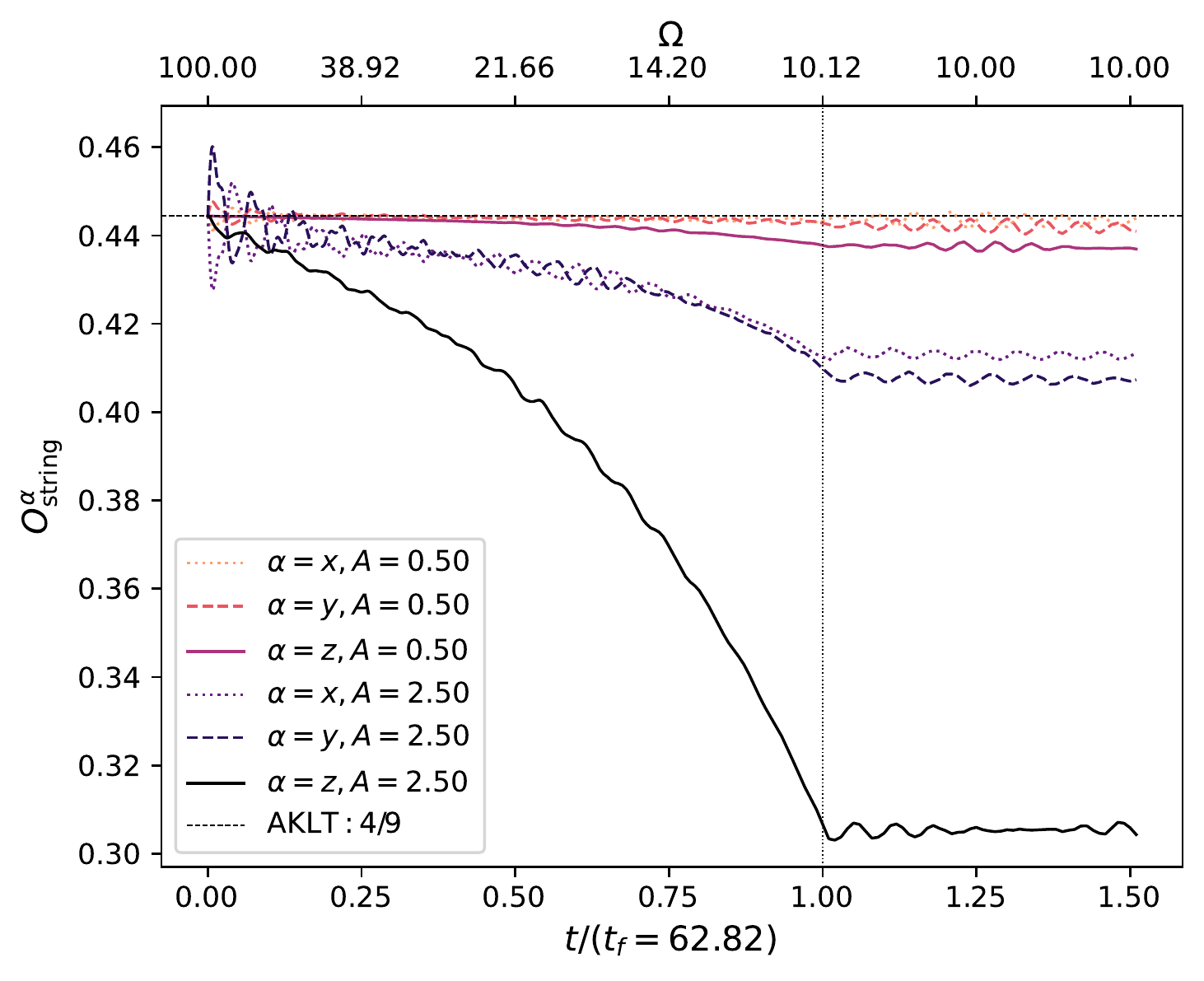}
\caption{Time evolution of the string order parameter Eq.~\eqref{string_order_parameter} in the $x,y,z$ directions during the drive starting from the Haldane phase (see Sec.~\ref{sec:Haldane}) for $\Omega_{f}=10.0$, $A=0.5,2.5$. The drive fails to suppress the string order, with the transverse $z$-direction being the most affected. The initial AKLT value of $4/9$ is marked by the horizontal dashed line. The Haldane phase is therefore robust against the drive Eq.~\eqref{protocol_XY}.}
\label{figure_2}
\end{figure}

We apply the protocol given by Eqs.~\eqref{ham_quench} and~\eqref{protocol_XY} to the AKLT state, which is the ground state of the Hamiltonian Eq.~\eqref{aklt_hamiltonian} ($H_0=H_{\text{AKLT}}$) and an ideal representative of the Haldane phase.

Since the drive does not break the protective symmetries of the Haldane phase, we expect the phase to be robust at least against small amplitudes. Numerically, we are able to go up to $A=2.5$ and find that the drive is still unable to destroy the Haldane phase because it is unable to destroy the string order. For the initial AKLT state, the string order is isotropic with a value of $O_{\text{string}}^{\alpha=x,y,z}=4/9$. Figure~\ref{figure_2} shows its evolution and we find that it remains finite in all directions. (Note that this stands in contrast with a sudden quench, which is able to destroy the order~\cite{Mazza_2014,CalvaneseStrinati_2016}.) Furthermore, we find no noticeable increase of a quadrupole polarization or correlations (not shown).

Our conclusion is that the robustness of the Haldane phase is difficult to overcome in our setup and one needs to start from an initial ground state which should already incorporate the breaking of its protective symmetries. To this end, we explore the driving protocol of Eq.~\eqref{protocol_XY} and apply it to the initial Hamiltonian given by Eq.~\eqref{Heisenberg_hamiltonian} with non-zero $h$ and $D$.

\begin{figure}[t!]
\includegraphics[scale=0.58]{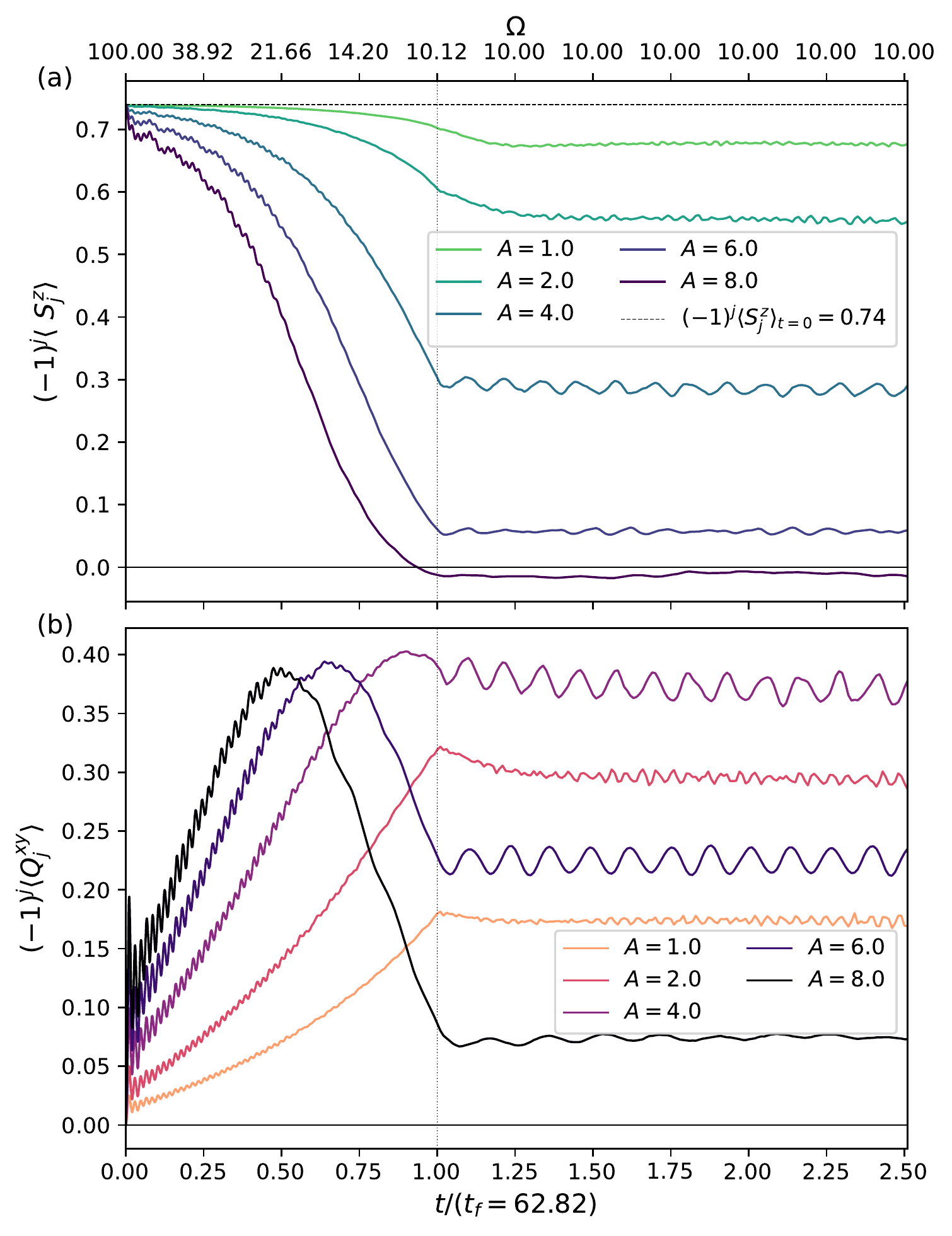}
\caption{
\label{figure_5}Time evolution after a quench starting from the trivial N\'eel phase ($\Omega_f=10$) of (a) the staggered magnetization $(-1)^j\avg{S^z_j}$, decaying as a function of time for strong values of the drive amplitude $A$ [Eq.~\eqref{protocol_XY}], and (b) the net quadrupole moment $|(-1)^{j}\langle Q_{j}^{xy}\rangle|\neq 0$ [see Eq.~\eqref{eq:Q}] induced by the drive. Both observables remain steady for times $t>t_{f}$. }
\end{figure}

\begin{figure*}[t!]
\begin{center}
    \includegraphics[scale=0.58]{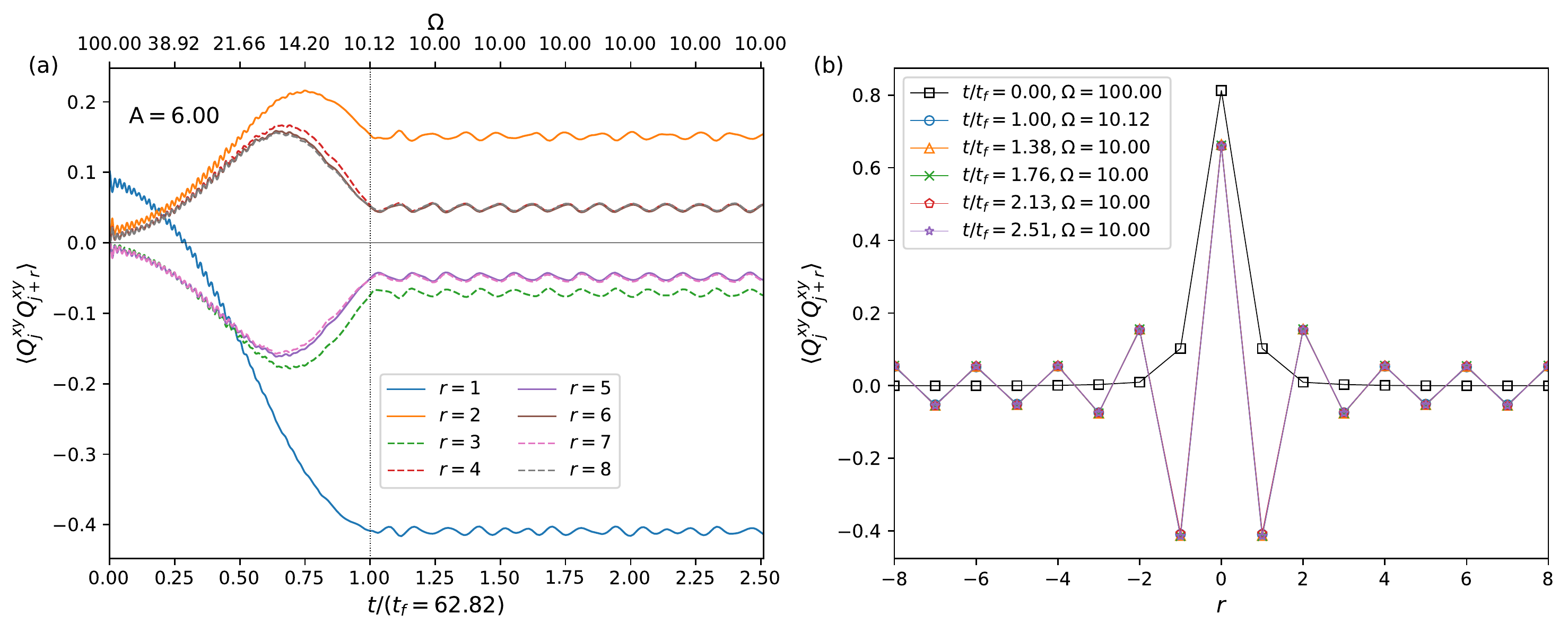}
    \caption{(a) Time evolution for the quadrupole correlations $\langle Q_{j}^{xy}Q_{j+r}^{xy}\rangle$ [see Eq.~\eqref{eq:Q}] for the setup as in Fig.~\ref{figure_5} and an amplitude of the drive $A=6.0$ [Eq.~\eqref{protocol_XY}]. We observe the development of an antiferroquadrupolar (AFQ) state, in which odd (even) values of $r$  become negative (positive). (b) The corresponding AFQ profile measured at specific time points $t/t_{f}\geq 1$, compared to the initial value at $t=0$; straight lines between data points are included as a guide to the eye. The overlap of profiles at constant $\Omega=\Omega_{f}$ for times $t/t_{f}\geq 1$ indicates the survival of the AFQ state in the steady-state limit.}
    \label{figure_6}    
\end{center}
\end{figure*}

\subsection{Driving from a trivial phase}
\label{subsec:driving_trivial_phase}

\subsubsection{Choice of parameters}

Relatively small anisotropies and interchain couplings are enough to induce an ordered phase for the $S=1$ chain. Still, in a lot of Ni-based compounds both are weak enough to keep them in the Haldane phase or at the edge of the phase transition line~\cite{Wierschem_2014}. An exception is CsNiCl$_3$, which shows Néel order below a critical temperature, and for which $h=0.051$, $D=-0.038$ have been deduced~\cite{Morra_1988,Sakai_Takahashi_1990} (though not without uncertainty~\cite{Buyers_1986}). 
A very large anisotropy of $D\sim-1.5$ was recently proposed for BaMo(PO$_4$)$_2$~\cite{Abdeldaim2019}.

To demonstrate the general principle, we set $H_{0}=H_{\text{Heis}}$ [Eq.~\eqref{Heisenberg_hamiltonian}], $D=-0.2$ and $h=+0.1$, which puts the initial system firmly into the ordered Néel phase: The ground state has a finite staggered spin polarization of $(-1)^j\avg{S^z_j}\approx0.74$ in the $z$-direction. Spin-spin correlations of the $z$-projection show antiferromagnetic (AFM) long-range order, while quadrupolar correlations are short-ranged (see Fig.~\ref{figure_6}(b) for $t=0$).

\subsubsection{\label{subsubsec:afq_pol}Suppression of AFM Ne{\'e}l order and emergence of quadrupole polarization}  
The time evolution of the staggered magnetization $(-1)^j\avg{S^z_j}$ and the quadrupole component $(-1)^j\avg{Q^{xy}_j}$ are presented in Fig.~\ref{figure_5}. The initial staggered magnetization of the N{\'e}el state is always reduced by the drive and is completely suppressed for $A\approx8$. This coincides with the emergence of a staggered net quadrupole moment in the $Q^{xy}$-direction, which becomes finite and survives in the steady-state limit, where $\Omega_f=10.0$ is kept constant. There is an optimal value of $A$ where the quadrupole polarization is highest, i.e., it tends to be smaller for both very small and very large $A$.

We note that the polarization is obtained in the third component $Q^{xy}$ of the quadrupolar operator Eq.~\eqref{eq:Q}, while the driving term only couples to the first component $Q^{x^{2}-y^{2}}$. This axial enhancement is somewhat analogous to the behaviour observed in Ref.~\onlinecite{Takayoshi}, where switching on a rotating magnetic field in the $xy$-plane induces a finite polarization of the $z$-component of the spin, i.e., perpendicular to the plane of the driving.

\subsubsection{\label{subsubsec:afq_corr}The antiferroquadrupolar correlations}
The above result suggests the dynamical emergence of a spin-nematic state due to application of the drive, where the quadrupole polarization dominates over the magnetization. To further investigate this state, we fix $A=6.0$ and show the quadrupole correlations $\langle Q_{j}^{xy}Q_{j+r}^{xy}\rangle$ in Fig.~\ref{figure_6}. The appearance of an antiferroquadrupolar (AFQ) state is characterized by negative (positive) correlations at odd (even) sites $r$ that remain stable for times $t/t_{f}>1$ in the steady-state limit. Note that there is a sign switch as a function of time for $r=1$. In Fig.~\ref{figure_6}(b), we show the profile of the quadrupole correlations characteristic of the AFQ state in the steady-state limit, compared to their initial values at $t=0$.

We point out that our resulting state is distinct from the large-$E$ phases reported in Ref.~\onlinecite{Tzeng_2017}, where the polarization is in the direction of $Q^{x^2-y^2}$, i.e. along the applied field $E$. If the ground state can be chosen to be real-valued, then $\avg{Q^{xy}_j}=0$ must necessarily hold in equilibrium. A polarization in the direction of $Q^{xy}$ therefore results from the time propagation during the drive, which gives an imaginary part to the wavefunction.
Thus, our resulting state generally has no equilibrium analogue for systems described by Eqs.~\eqref{Heisenberg_hamiltonian} and \eqref{bi_quad}, even if the rhombic anisotropy term Eq.~\eqref{eq:rhombic_an} is also present, unless spontaneous breaking of time-reversal symmetry results in a complex-valued ground state. We discuss this in some more detail in Sec.~\ref{sec:AFQwavefunction}.

\begin{figure*}[t!]
\begin{center}
    \includegraphics[scale=0.54]{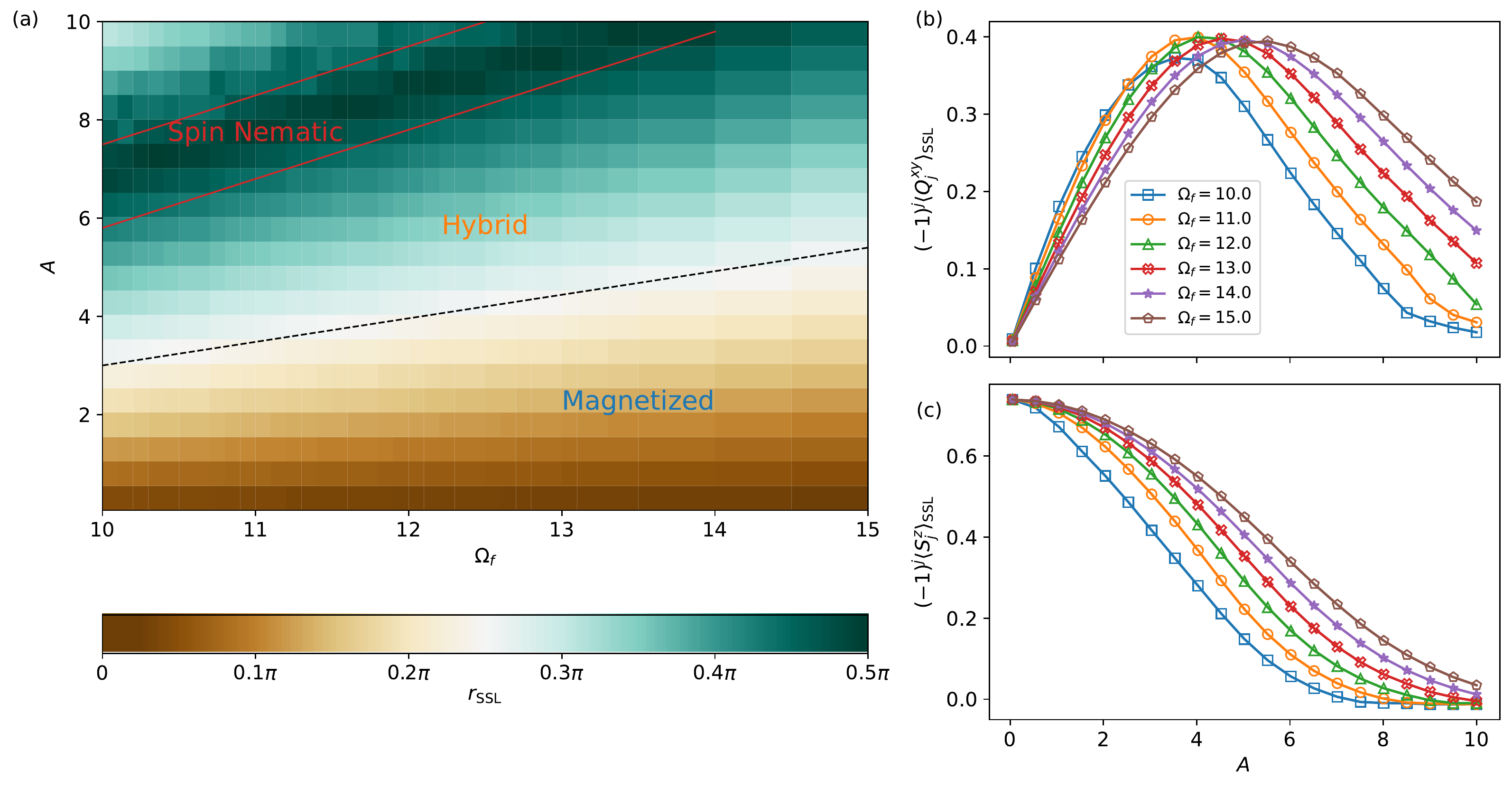}
    \caption{(a) Phase diagram in the steady-state limit (SSL) following a quench from the trivial N\'eel phase for the crossover parameter $r_{\text{SSL}}$ defined in Eq.~\eqref{eq:r_ssl_def}. The three regimes are (1) magnetized: $r_{\text{SSL}}<\pi/4$, (2) hybrid: $r_{\text{SSL}}\gtrapprox \pi/4$, (3) spin-nematic: $r_{\text{SSL}}\approx \pi/2$. (b) The SSL values of the emerging quadrupole moment for different $\Omega_{f}$ as a function of the driving amplitude $A$. Reaching smaller $\Omega_{f}$ enhances the net quadrupole moment requiring smaller amplitudes $A$. Strong values of $A$ tend to suppress the overall enhancement of the quadrupole moment, cf. Fig.~\ref{figure_5}. (c) The SSL value for the staggered magnetization at different $\Omega_{f}$ as a function of the driving amplitude $A$. A smaller $\Omega_{f}$ requires smaller $A$ to suppress the Ne{\'e}l order. Straight lines between data points in (b) and (c) are a guide to the eye.
    }
    \label{figure_7}    
\end{center}
\end{figure*}

\subsubsection{\label{subsubsec: afq_pd }Phase diagram in the steady-state limit}

To better characterize the dynamically obtained AFQ state and explore its stability at longer times, we study the phase diagram in the steady-state limit (SSL) for the two order parameters of the staggered magnetization $(-1)^{j}\langle S_{j}^{z}\rangle$ and the quadrupole polarization $(-1)^{j}\langle Q_{j}^{xy}\rangle$, as a function of the target frequencies and driving amplitudes $(\Omega_{f},A)$. The spin-nematic regime is reached when $(-1)^{j}\langle S_{j}^{z}\rangle_{\text{SSL}}\approx0$ and $(-1)^{j}\langle Q_{j}^{xy}\rangle_{\text{SSL}}\neq 0$ for an average over the steady-state regime Eq.~\eqref{eq:ssl_def}. We define the crossover parameter between the two regimes as:
\begin{eqnarray}\label{eq:r_ssl_def}
r_{\text{SSL}}=\arctan\left(\left|\frac{\langle Q_{j}^{xy}\rangle_{\text{SSL}}}{\langle S_{j}^{z}\rangle_{\text{SSL}}}\right|\right)\in\left[0,\frac{\pi}{2}\right].
\end{eqnarray}

Figure~\ref{figure_7}(a) shows the phase diagram for $r_{\text{SSL}}$. We distinguish three well-differentiated regions: There is a \textit{magnetized} region $r_{\text{SSL}}\approx 0$, where the staggered magnetization dominates, as in the initial ground state. The black dashed line is an orientative contour line for $r_{\text{SSL}}=\pi/4$, which separates the magnetized region from a region where the two polarizations are of approximately equal strength, which we call the \textit{hybrid} region. Finally, the straight lines delineate the \textit{spin-nematic} region $r_{\text{SSL}}\approx \pi/2$, where $\langle S^{z}_{j}\rangle\to 0$.

The effect of reducing the target frequency $\Omega_{f}$ on both the staggered magnetization and the quadrupolar moment is shown in Figs.~\ref{figure_7} (b), (c). 
In the small-$A$ region, the systematics of the AFM order and AFQ order are opposite, so that decreasing the frequency leads to weaker AFM polarization and stronger AFQ polarization. In the large-$A$ region, the systematics is the same and decreasing the frequency leads to a weakening in both.

\subsubsection{\label{sec:AFQwavefunction}The AFQ wavefunction}

Despite the apparent simplicity of the final xy-AFQ state, we find that it is entangled and cannot be written down as a simple analytical wavefunction. However, we can heuristically attempt to write it down as a simplified MPS by restricting ourselves to the three largest eigenvalues of the entanglement spectrum $\Lambda^{\text{AB}}$ and $\Lambda^{\text{BA}}$ (for the sublattices A and B) in Eq.~\eqref{eq:mps_laga}.

Analyzing the numerical result for the ground state, we find that it approximately has the following MPS structure:
\begin{equation}\label{eq:initial_3_mps}
\begin{split}
\Gamma^{\text{A},+} &=\left(\begin{matrix}
-a_{11}^{+} & 0 & 0\\
0 & 0& 0\\
0 &0&0
\end{matrix}\right),\\
\Gamma^{\text{A},0} &=\left(\begin{matrix}
0 & a_{12}^{0} & 0\\
-a_{12}^{0} & 0& 0\\
0 &0&0
\end{matrix}\right),\\
\Gamma^{\text{A},-} &=\left(\begin{matrix}
0 & 0 & -a_{13}^{-}\\
0 & a_{22}^{-}& 0\\
a_{13}^{-} &0&0
\end{matrix}\right),\\
\Gamma^{\text{B},+} &= -\Gamma^{\text{A},-}\\
\Gamma^{\text{B},0} &=-\Gamma^{\text{A},0},\\
\Gamma^{\text{B},-}&=-\Gamma^{\text{A},+},
\end{split}
\end{equation}
where $a_{ij}^{\sigma}=\left(\Gamma^{A,\sigma}\right)_{ij}$ denotes the non-zero matrix entries.

After switching on the drive, we keep monitoring the pattern of non-zero entries and find that the xy-AFQ state shows the following structure:
\begin{equation}\label{eq:mps_afmq}
\begin{split}
\Gamma^{\text{A},+}[a_{ij}^{+}(t)] &\approx\left(\begin{matrix}
a_{11}^{+}(t) & 0 & a_{13}^{+}(t)\\
0 & a_{22}^{+}(t)& 0\\
a_{31}^{+}(t) &0&a_{33}^{+}(t)
\end{matrix}\right),\\
\Gamma^{\text{A},0}[a_{ij}^{0}(t)] &\approx\left(\begin{matrix}
0 & a_{12}^{0}(t) & 0\\
a_{21}^{0}(t) & 0& a_{23}^{0}(t)\\
0 &a_{32}^{0}(t)&a_{33}(t)
\end{matrix}\right),\\
\Gamma^{\text{A},-}[a_{ij}^{-}(t)] &\approx\left(\begin{matrix}
a_{11}^{-}(t) & 0 & a_{13}^{-}(t)\\
0 & a_{22}^{-}(t)& 0\\
a_{31}^{-}(t) &0&a_{33}^{-}(t)
\end{matrix}\right),\\
\Gamma^{\text{B},\sigma}(t) &= \Gamma^{\text{A},\sigma}[a_{ij}^{\sigma}(t)\to b_{ij}^{\sigma}(t)],\hspace{10pt}\sigma\in\{+,0,-\},\\
\Lambda^{\text{AB}}(t)&= \text{diag}\left[\Lambda_{0}^{\text{AB}}(t),\Lambda_{1}^{\text{AB}}(t),\Lambda_{2}^{\text{AB}}(t)\right],\\
\Lambda^{\text{BA}}(t)&=\Lambda^{\text{AB}}(t).
\end{split}
\end{equation}
Moreover, we also observe:
\begin{eqnarray}
|a_{ij}^{\pm}(t)|\approx |b_{ij}^{\mp}(t)|, \hspace{10pt} |a_{ij}^{0}(t)|\approx |b_{ij}^{0}(t)|.
\end{eqnarray}
As a technical detail, note that the reduced MPS also needs to be renormalized.

\begin{figure}[t!]
\includegraphics[scale=0.58]{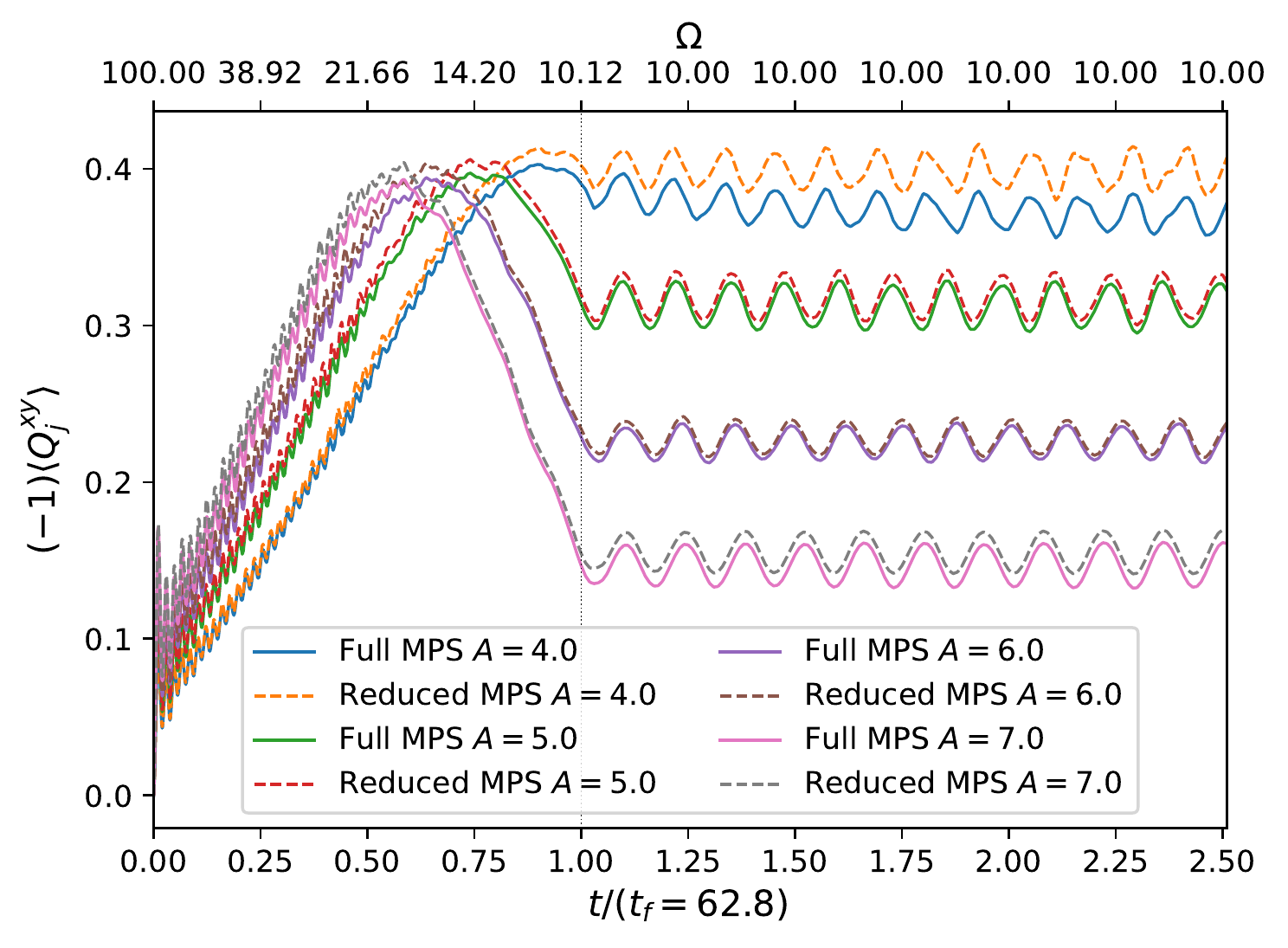}
\caption{Comparison between the net quadrupole moment [Eq.~\eqref{eq:Q}] in the $Q^{xy}$-direction obtained from the full MPS; and the reduced MPS obtained by truncating to the three dominant eigenvalues of the reduced density matrix (see Sec.~\ref{sec:AFQwavefunction}). The qualitative behaviour is reliably captured by the smaller MPS.}
\label{figure_8}
\end{figure}

Figure~\ref{figure_8} compares $\langle Q_{s=\text{A}, \text{B}}^{xy}\rangle$ obtained with the full and with the heuristically reduced wavefunction, with overall good agreement. To derive an analytic expression for $\langle Q_{s=\text{A}, \text{B}}^{xy}\rangle$, we note that $Q_{s}^{xy}$ can be written using $Q^{\pm}_s=\left(S^{\pm}_s\right)^2$, $S^{\pm}_s=S^x_s\pm\mathrm{i}S^y_s$ as follows:
\begin{equation}\label{eq:q_xy_projectors}
\begin{split}
Q_{s}^{xy} &= \mathrm{i}/2 \left(Q_{s}^{-}-Q_{s}^{+}\right),\\
\avg{Q^{xy}_s} &= \text{Im}\left(\avg{Q^{+}_s}-\avg{Q^{-}_s}\right)/2.
\end{split}
\end{equation}
We find:
\begin{eqnarray}\label{eq:q_xy_def}
\langle Q_{s}^{\pm}\rangle =  \frac{1}{2}\sum_{i,j}\left[\Lambda_{i}^{\text{AB}}\Lambda_{j}^{\text{BA}}\right]^{2}\left(\Gamma_{ji}^{s,\pm}\right)^{*}\Gamma_{ij}^{s,\mp}.
\end{eqnarray}

Equation~\eqref{eq:q_xy_def} shows that for the initial MPS in Eq.~\eqref{eq:initial_3_mps}, $\langle Q_{s}^{\pm}\rangle=0$, as the matrices $\Gamma^{s,\pm}$ do not share common matrix elements at equal row and column indices. Conversely, the driven state given by Eq.~\eqref{eq:mps_afmq} does contain non-zero products between the matrix elements and yields a net quadrupole moment $\avg{Q^{xy}_j}$.

Furthermore, we see that any contribution to $\avg{Q_{s}^{xy}}$ must come from the imaginary part of $\avg{ Q^{\pm}_{s}}$. This implies that the matrix elements of $\Gamma^{s,\pm}$ must contain imaginary parts, which are acquired in the course of the unitary time evolution (beyond a trivial global phase). This shows that the obtained xy-AFQ state cannot be engineered from an equilibrium configuration by means of an arbitrary variation of the parameters of the model, provided that time-reversal symmetry holds (i.e., the ground-state wavefunction can be chosen real-valued).

\section{Summary}

We have demonstrated that it is possible to convert a conventional N\'eel antiferromagnetic state on the $S=1$ quantum spin chain into an unconventional antiferroquadrupolar state by applying the adiabatic Floquet protocol (starting with an $\Omega=\infty$ initial state and adiabatically decreasing $\Omega$). We chose to drive the rhombic anisotropy Eq.~\eqref{protocol_XY}, as it contains squares of the spin operators. Up to the time scales considered, the engineered state remains stable when the final frequency is kept fixed. The range of amplitudes $A$ and final frequencies $\Omega_f$ where such an AFQ state is observed is shown in the phase diagram of Fig.~\ref{figure_7}. While we are limited to relatively large $\Omega_f$ in terms of numerics, we surmise that the spin-nematic region extends to smaller $\Omega_f$ (and therefore smaller $A$ according to Fig.~\ref{figure_7}).

In contrast to this, we find that an initial state belonging to the symmetry-protected Haldane phase is robust against such driving, as evidenced by the preserved string order parameter. Thus, while a lot of the previous interest in $S=1$ chains was motivated by an experimental realization of Haldane physics, we find that its robustness limits the possibilities for nonequilibrium engineering. We therefore propose that conventional Néel-ordered compounds are more useful in this regard. In particular, we see that the easy-axis anisotropy is crucial in stabilizing a net quadrupole moment.

Overall, the adiabatic Floquet protocol presents a controlled approach to driven systems in order to manipulate and engineer valuable quantum states. Due to the adiabatic changes, we can achieve much longer numerical propagation times than in the case of sudden quenches.

From a technical perspective, we have employed matrix product states in combination with the iTEBD algorithm; we have also used exact diagonalization (see App.~\ref{app:tests}) to back up our claims. Due to the adiabatic nature of the quench, a bond dimension of $\chi=200$ is sufficient to bound the discarded weight from above. In fact, we provided analytical arguments that the drive from the N\'eel state can essentially be encoded with $\chi=3$. For small discarded weights shown in App.~\ref{app:tests}, however, we observe a fast increase of the bond dimension and thus in principle an uncontrolled error on this scale. This might be attributed to non-generic features of the micromotion, though we lack a clear understanding of this issue, which is beyond the scope of this paper.

An experimental realization for the driven $S=1$ spin chain considered here can be either using lasers with real materials~\cite{Takayoshi,Seifert2022,Lemmens_2021} (cf. Fig.~\ref{fig:figure_model}), or in cold-atom systems~\cite{Senko_2015}.

\section*{Acknowledgements}

We acknowledge support by the `Nieders\"achsisches Vorab' through the `Quantum- and Nano-Metrology (QUANOMET)' within  the  project  P-1. The authors would like to thank the Publication Fund of the TU Braunschweig for financially supporting the freely accessible publication of this article.

\appendix

\section{\label{app:time_discretization}Details of the time discretization}

To perform an adiabatic variation of the driving frequency $\Omega$ we proceed as follows: We logarithmically discretize~\footnote{The choice of a logarithmically spaced grid for the interval $\left[\Omega_{f},\Omega_0\right]$ is just for computational purposes, favouring smaller period increments as we approach $t_{f}$. In a purely adiabatic process, such discretization choice should be irrelevant.} the interval $\left[\Omega_{f},\Omega_{0}\right]$ into $N_{\text{cycles}}+1$ points $\left\{\Omega_{j}\right\}_{j=0,...,N_{\text{cycles}}}$, where $N_{\text{cycles}}$ determines the total number of cycles completed by the drive in the time window $\left[0,t_{f}\right]$; thus, $t_{f}$ is completely determined by the $N_{\text{cycles}}$ parameter.

The adiabatic limit is realized with $N_{\text{cycles}}\to\infty$ (equivalently $t_{f}\to\infty$).
Numerically, we fix $N_{\text{cycles}}=300$. After a single cycle is performed with period $T_j$ from times $\left[t_{j},t_{j+1}\right]\in[0,t_{f}]$, we decrease the frequency by $\delta\Omega_{j}$ from $\Omega_j\to\Omega_j-\delta\Omega_{j}$, with the corresponding increment in the period $T_j\to T_j+\delta T_j$. For times $t>t_{f}$, we keep the frequency constant and equal to $\Omega_{f}$, evolving the state to a final time $t_{\text{end}}=\alpha t_{f}$.

\section{\label{app:tests}Numerical tests}

\begin{figure}[b]
\includegraphics[scale=0.56]{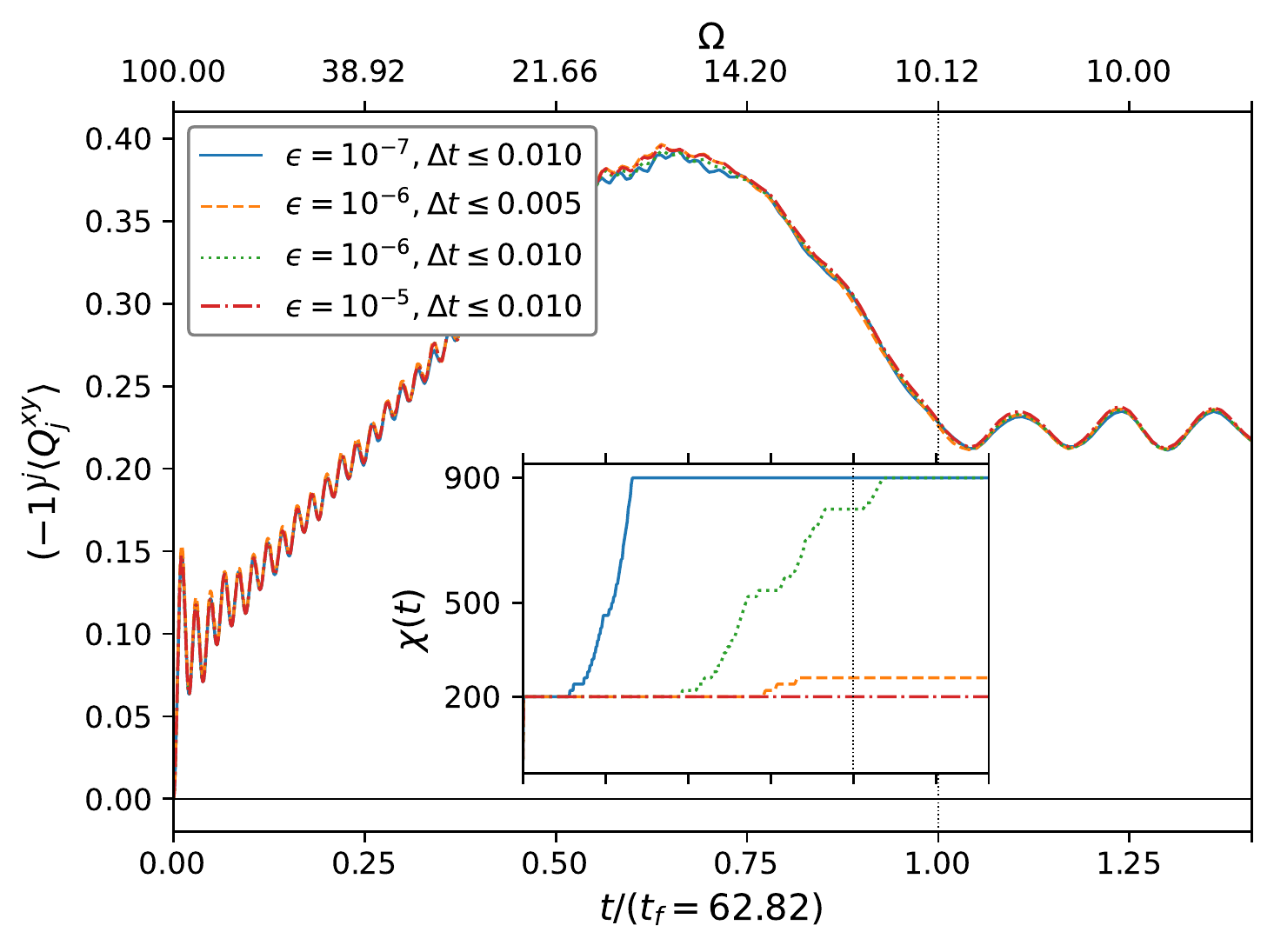}
\caption{The same as Fig.~\ref{figure_5}(b) with $A=6$ but for different discarded weights $\epsilon$ and maximum Trotter steps $\Delta t$. \emph{Inset}: The corresponding MPS bond dimension as a function of time. Horizontal axis ticks and labels correspond to those of the main panel.  }
\label{figure_9}
\end{figure}

In Fig.~\ref{figure_9}, we show the quadrupole polarization $(-1)^{j}\langle Q_{j}^{xy}\rangle$ following a quench from the N\'eel phase for different values of the discarded weight $\epsilon$ and maximum Trotter steps $\Delta t$ (the setup is analogous to Fig.~\ref{figure_5}(b) which was obtained using $\epsilon=10^{-5}$ and $\Delta t=0.01$). The corresponding bond dimension is shown in the inset. Once a maximum value of $\chi=900$ has been reached, the bond dimension is no longer increased due to computational limitations, and the time evolution is carried out using a fixed $\chi$. We note that the error is no longer strictly controlled in this regime.

At a maximum $\Delta t=0.01$, the bond dimension does not increase for $\epsilon=10^{-5}$, but it increases mildly for $\epsilon=10^{-6}$ and rapidly for $\epsilon=10^{-7}$, where its maximum value of $\chi=900$ is reached quickly and the error is no longer controlled. This issue persists even if the quench is made more adiabatic ($N_\textnormal{cycles}=500$). Physical quantities, such as the quadrupole polarization, however, seem converged. It is reasonable to assume that our adiabatic protocol can be modelled accurately using a small $\chi$ (in fact, $\chi=3$ seems sufficient, see the analytic discussion in Sec.~\ref{sec:AFQwavefunction}). A possible explanation is that the blow-up of the bond dimension at small $\epsilon$ is related to features of the micromotion.

\begin{figure}[t]
\includegraphics[scale=0.48]{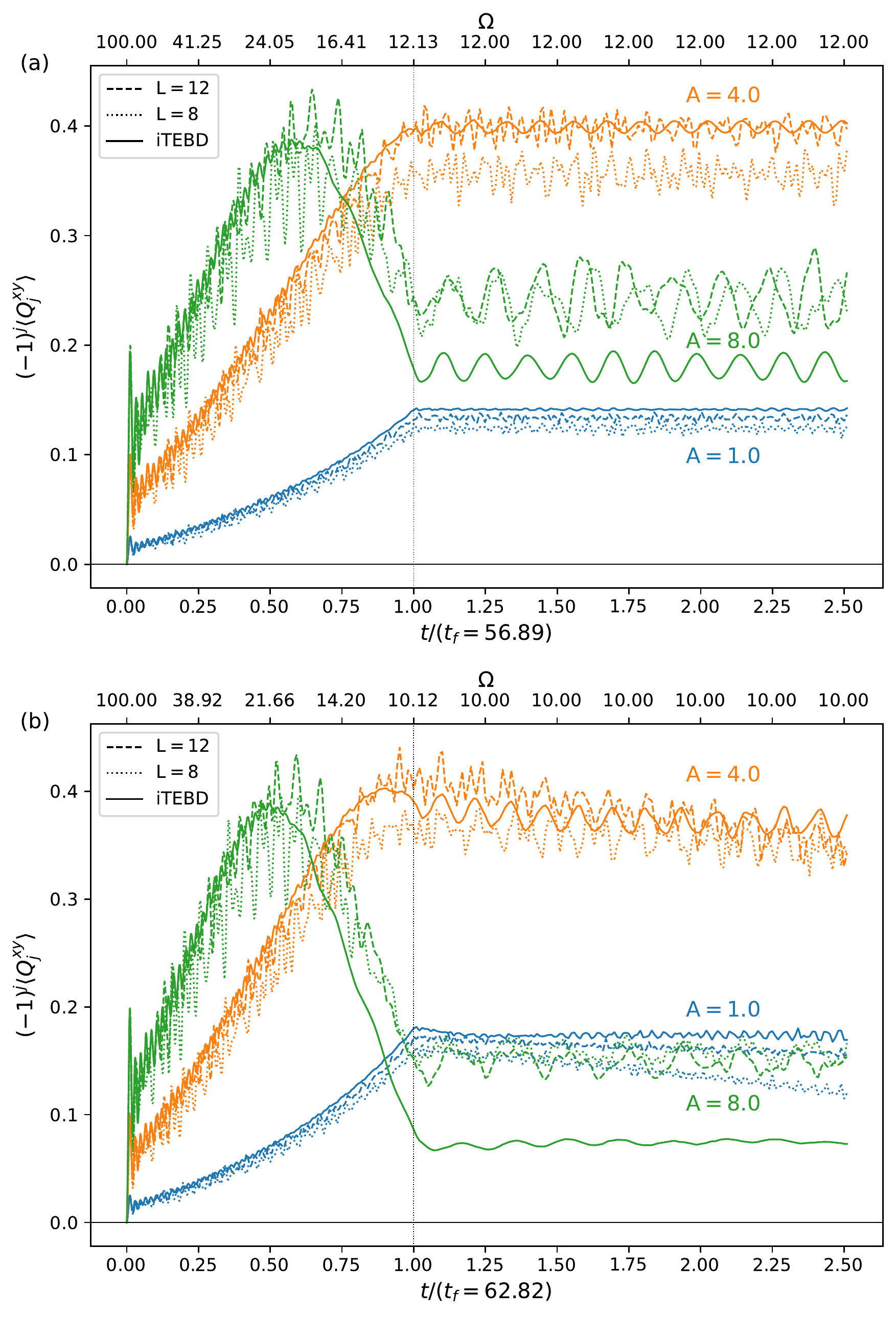}
\caption{Comparison of iTEBD and ED results for the time evolution of the net quadrupole moment following a quench starting from the N\'eel phase for (a) $\Omega_{f}=12.0$, and (b) $\Omega_{f}=10.0$ (the latter corresponds to Figs.~\ref{figure_5} and \ref{figure_9}).}
\label{figure_10}
\end{figure}

We add supportive evidence by comparing with exact diagonalization (ED) data obtained system sizes of $L\leq12$ with periodic boundary conditions. The results are shown in Fig.~\ref{figure_10}. One can see that both approaches agree qualitatively and even quantitatively unless $\Omega_f$ is small or $A$ is large. In particular, the ED data indicates a finite steady-state value of the quadrupole polarization (or a trend in this direction as $L$ is increased).

\begin{figure}[t]
\includegraphics[scale=0.55]{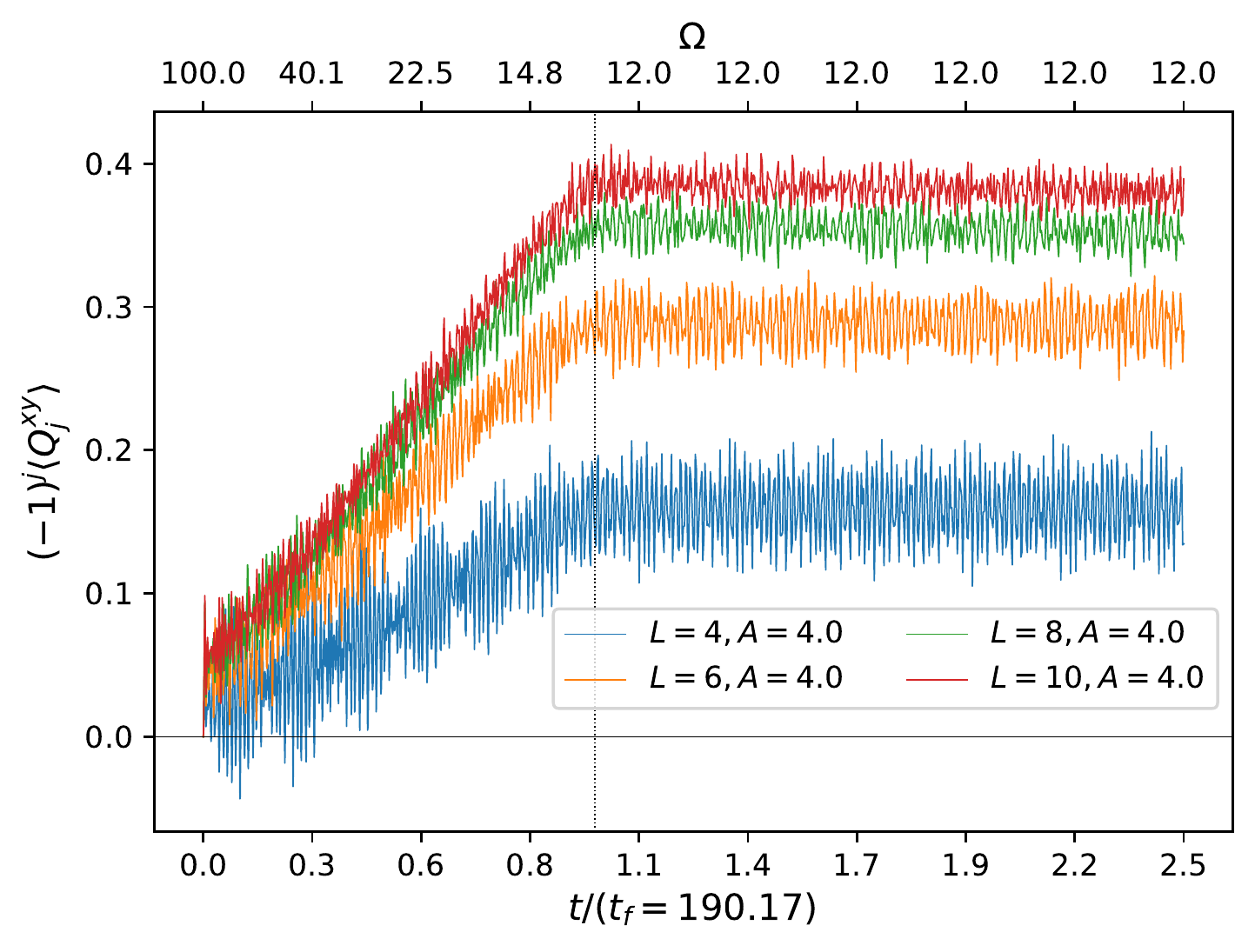}
\caption{Study of the stability of the quasi-steady state for $A=4$, $\Omega_f=12$ and very long propagation times ($N_{\text{cycles}}=1000$) using exact diagonalization in a small system of $L=4,6,8,10$.}
\label{figure_11}
\end{figure}

Finally, since time propagation using ED poses no entanglement problem, we check the robustness of the quasi-steady state for very long propagation times, using $N_{\text{cycles}}=1000$, so that $t_f\approx190$, $t_{\text{end}}=2.5t_f\approx475$ and finite systems of up to $L=10$. The result is shown in Fig.~\ref{figure_11}. We observe that the quasi-steady state remains robust in this parameter range.

\section{\label{app:Floquet Hamiltonian in adiabatic evolution}}
In this section, we present general expressions for the Floquet adiabatic evolution that can be used as a starting point for methods beyond the direct time propagation used in this paper.

Given a purely periodic Hamiltonian $H(t+T)=H(t)$ with period $T$, the Floquet Hamiltonian $H_F$ is formally defined by the unitary operator over a single cycle in the time interval $[t_0,t_0+T]$:
\begin{eqnarray}\label{eq:floq_un}
U(t_0+T,t_0)=\hat{T}e^{-i\int_{t_0}^{t_0+T}ds H(s)}\equiv e^{-iH_F T},
\end{eqnarray}
where $\hat{T}$ is the time-ordering operator. Note that in general, the Floquet Hamiltonian is dependent both $t_0$, and $T$; i.e. $H_F\equiv H_F(t_0,T)$. In most of cases determining $H_F$ in an exact way is not possible due to the complex time ordering appearing on the right hand side. A common approach to approximate $H_F$ for a fixed frequency $\Omega=2\pi/T$ is to truncate the Magnus expansion to a given order or to employ an effective Hamiltonian; however, the validity of these approaches is normally restricted to the region of high $\Omega$ values. 

In the adiabatic Floquet approach described in Sec.~\ref{subsec:adiabatic_protocol}, we start with the initial frequency $\Omega_0=\infty$, where $H_F$ is known exactly. Due to the slow decrease in $\Omega$, $H_F$ undergoes infinitesimal changes (and so does the unitary evolution operator in Eq.~\eqref{eq:floq_un}). To determine the adiabatic variation of $H_F$, we look at two neighbouring cycles with the drive period differing by an infinitesimal amount $\delta T$, and employ in their respective intervals the definition given by Eq.~\eqref{eq:floq_un}. This way, one does not need to solve any of the Floquet Hamiltonian problems individually, as only the difference between the series expansions of Eq.~\eqref{eq:floq_un} needs to be taken into account.   

We discretize the time interval $[t_0,t_{f}]$ as a set of points $t_{k=0,...,N=f}$. The values of $t_k$ will be called the \textit{switching} times, representing the times where an infinitesimal variation of the driving period $\delta T$ takes place. In each interval $[t_{k-1},t_{k-1}+T_{k}]$, the value of the drive frequency is fixed and given by $\Omega_{k}=2\pi/T_{k}$, and the drive function $f(t)$ satisfies:
\begin{eqnarray}
f(t+T_k)=f(t) \hspace{5pt}\forall t\in\left[t_{k-1},t_{k-1}+T_{k}\right].
\end{eqnarray}

We consider a time dependent Hamiltonian of the form:
\begin{eqnarray}\label{eq:eq_c1}
H(t)&=&H_0 + V(t),\nonumber\\
V(t)&=&f(t)\sum_{j}X_{j} = f(t)X,
\end{eqnarray} 
where $H_0$ is the time-independent part and $X_j$ are generic local operators. To simplify calculations, we further impose the following restriction on the drive function within each finite interval:
\begin{eqnarray}\label{eq:ft_cons}
f(0)&=&f(T)=0,\hspace{10pt}\int_{t_{k-1}}^{t_{k-1}+T_{k}}ds~f(s)=0,\hspace{5pt}\forall k.
\end{eqnarray}

Since the initial Hamiltonian $H_0$ is known, one can choose to start from an eigenstate of $H_0$. In that case, it is more convenient to work in the interaction representation, where the time evolution operator is explicitly given by:
\begin{eqnarray}
U(t_f,t_{0})=\hat{T}e^{-i\int_{t_0}^{t_f}ds~ e^{iH_0 s}V(s)e^{-iH_0 s}}.
\end{eqnarray}
On each time interval $[t_{k-1},t_{k-1}+T_{k}]$, we associate a Hermitian operator $H_{F}^{(k)}$ defining the Floquet Hamiltonian in that interval:
\begin{eqnarray}
U(t_{k-1}+T_{k},t_{k-1})&\equiv& e^{-i H_{F}^{(k)}T_{k}}=U_{F}^{(k)},\nonumber\\
U(t_f,t_0)&=&\prod_{k=1}^{N}U_{F}^{(k)}.
\end{eqnarray}
The last identity follows from the connection property of the evolution operator. Within a given time interval of fixed period $T_k$, the series expansion for the unitary operator is:
\begin{eqnarray}
U_{F}^{(k)}=\sum_{n=0}^{\infty}\frac{(-i)^{n}}{n!}\int_{t_{k-1}}^{t_{k-1}+T_{k}}\prod_{j=1}^{n}dt_{j}\hat{T}\left[\hat{V}_k(t_{j})\right],
\end{eqnarray}
where the carets indicate that the operators are in the interaction representation, i.e. $\hat{V}_k(s)=e^{iH_0 s}V_k(s)e^{-iH_0 s}$, and we define $V_{k}(s)\equiv f(s,T_k)X$ (see Eq.~\eqref{eq:eq_c1}) for a fixed value of the period $T_k$. The term in brackets contains a product of the operators $\hat{V}_k(t_j)$ whose order is irrelevant due to the time ordering operator acting over the bracket.

We consider now two adjacent unitary operators:
\begin{eqnarray}
U_{F}^{(k)}&=&e^{-iH_{F}^{(k)}T_{k}},\nonumber\\
U_{F}^{(k+1)}&=& e^{-iH_{F}^{(k)}(T_{k}+\delta T)-i\delta V_{k}(T_{k}+\delta T)},
\end{eqnarray}
with $\delta T\sim 0$ an infinitesimal variation of the period, and $\delta V_k$ representing an infinitesimal change of the Floquet operator when a decrease in the driving frequency from $\Omega_{k}$ to $\Omega_{k+1}$ takes place. The first interval of the evolution is $[t_{k-1},t_{k-1}+T_{k}]$ and the second is $[t_{k-1}+T_{k},t_{k-1}+2T_{k}+\delta T]$.

The initial form of the Floquet Hamiltonian is known in the $\Omega\to\infty$ limit:
\begin{eqnarray}
H_{F}^{(k=0)}=H_0.
\end{eqnarray}
The variation with respect to the period is given by:
\begin{eqnarray}
\lim_{\delta T\to 0}\frac{U_{F}^{(k+1)}-U_{F}^{(k)}}{\delta T}\equiv \partial_{T}U_{F}(T).
\end{eqnarray}

The two neighbouring terms are explicitly written in their series expansion:
\begin{eqnarray}
U_{F}^{(k)}&=& \sum_{n=0}^{\infty}\frac{(-i)^{n}}{n!}\int_{t_{k-1}}^{t_{k-1}+T_{k}}\prod_{j=1}^{n}dt_{j}\hat{T}\left[\hat{V}_{k}(t_{j})\right],\nonumber\\
U_{F}^{(k+1)}&=& \sum_{n=0}^{\infty}\frac{(-i)^{n}}{n!}\int_{t_{k-1}+T_{k}}^{t_{k-1}+2T_{k}+\delta T}\prod_{j=1}^{n}dt_{j}\hat{T}\left[\hat{V}_{k+1}(t_{j})\right].\nonumber\\
\end{eqnarray}
By changing the variables of integration in the $U_{F}^{k+1}$ to $\tilde{t}_{n}=t_{n}-T_{k}$, we obtain two different contributions when subtracting both series $\Delta U_{F}^{(k)}=U_{F}^{(k+1)}-U_F^{(k)}$:
\begin{eqnarray}\label{eq:Ufs}
\Delta U_{F}^{(k)}&=& \sum_{n=1}^{\infty}\frac{(-i)^{n}}{n!}\int_{t_{k-1}}^{t_{k-1}+T_{k}}\prod_{j=1}^{n}dt_{j}\hat{T}\left[\Delta\hat{O}_{k}(\{t_{j}\})\right] \nonumber\\
&+& \sum_{n=1}^{\infty}\frac{(-i)^{n}}{n!}\int_{t_{k-1}+T_{k}}^{t_{k-1}+T_{k}+\delta T}\prod_{j=1}^{n}dt_{j}\hat{T}\left[\hat{O}_{k+1}(\{t_{j}\})\right],\nonumber\\
\end{eqnarray}
where we have defined:
\begin{eqnarray}
\hat{O}_{k}(\{t_{j}\})&=&\hat{V}_{k}(t_{1})...\hat{V}_{k}(t_{n}),\nonumber\\
\Delta\hat{O}_{k}(\{t_{j}\})&=& \hat{O}_{k+1}(\{t_{j}\})-\hat{O}_{k}(\{t_{j}\}).
\end{eqnarray}
Note that all summed up terms start at $n=1$, since the identities appearing on the right hand side have cancelled out. For the $k+1$ term we make use of the following expansion:
\begin{eqnarray}
\hat{O}_{k+1}(\{t_{j}\})\approx \underbrace{\hat{V}_{k}(t_{1})...\hat{V}_{k}(t_{n})}_{=\hat{O}_{k}(\{t_{j}\})}+\delta T\sum_{j=1}^{n}\frac{\partial \hat{V}_{k}(t_{j})}{\partial T}\prod_{l\neq j}\hat{V}_{k}(t_{l}).\nonumber\\
\end{eqnarray}
We stress that the order of products inside the brackets is irrelevant due to the action of the time-ordering operator. The second term in the right-hand side of Eq.~\eqref{eq:Ufs} is identically zero when Eq.~\eqref{eq:ft_cons} is satisfied over the infinitesimal interval of integration. For the first series, the term in brackets at a given order $n$ is given by:
\begin{eqnarray}
(\delta T)\hat{T}\left[\sum_{j=1}^{n}\frac{1}{f(t_{j},T_k)}\left(\frac{\partial f(t_{j},T)}{\partial T}\right)\bigg|_{T=T_k}\hat{O}_{k}(\{t_{j}\})\right]\nonumber\\
\end{eqnarray}
Due to the time ordering, each summand will give exactly the same contribution to the propagator. Then we can select one of the intermediate times (say $t_1$), sweep it around the full string of operators, and write:
\begin{eqnarray}
&\alpha_n &\int_{a}^{b} dt_{1}w(t_1,T_k)\hat{V}_{k}(t_{1})\int_{a}^{b} dt_{2}...\int_{a}^{b} dt_{n}\hat{T}\hat{O}_{k}(\{t_{j}\}),\nonumber\\
\alpha_n &=& \frac{(-i)^{n}}{n!}(\delta T) n,
\end{eqnarray}
where we have abbreviated $a=t_{k-1}$ and $b=t_{k-1}+T_{k}$, and defined:
\begin{eqnarray}
w(t_1,T_k)=\left(\frac{\partial f(t_1,T)}{\partial T}\right)\bigg|_{T=T_k} \frac{1}{f\left(t_1,T_k\right)}.
\end{eqnarray}
 We can arrange $t_{1}$ a total of $n$ times, with the remaining $n-1$ operators still subjected to time-ordering. Proceeding to all orders, Eq.~\eqref{eq:Ufs} becomes:
\begin{eqnarray}\label{eq:uf_diff}
\Delta U_{F}^{(k)}\approx -i\delta T\int_{a}^{b}dt_{1}w(t_{1},T_k)\hat{V}_k(t_{1},T_k)U_{F}^{k}.
\end{eqnarray}
Eq.~\eqref{eq:uf_diff} gives a differential equation for the Floquet unitary operator. The initial condition is given by:
\begin{eqnarray}
U_{F}(T=0)=\lim_{T\to 0}e^{-iH_0 T}=1.
\end{eqnarray}
The corresponding Floquet Hamiltonian is then identified by the relation:
\begin{eqnarray}
U_{F}(T)&=&e^{-iH_{F}(T)T},\nonumber\\
H_{F}(T)&=&\frac{1}{T}\int_{t_{0}(T)}^{t_0(T)+T}ds e^{iH_0 s}V(s,T)e^{-iH_0 s}.
\end{eqnarray}
Note that the exact form of increments in the period $T$ is what determines the functional form of the switching times $t_0(T)$.

\FloatBarrier


%

\end{document}